\documentclass[11pt,letterpaper,notoc]{JHEP3}
\usepackage{amsmath,amsfonts,cite,epsfig}
\usepackage{feynmf}

\newcommand{\nn}{\nonumber}
\newcommand{\mpl}{M_{\rm Pl}}

\newcommand{\mv}{M_{\rm 5D}}

\newcommand{\be}{\begin{eqnarray}}
\newcommand{\ee}{\end{eqnarray}}
\newcommand{\hmn}{h_{\mu\nu}}
\newcommand{\cL}{{\cal L}}
\newcommand{\LFP}{{\cal L}_{\rm F P}}

\newcommand{\cA}{ {\cal A} }
\newcommand{\Lmin}{ {\cal L}_{\rm min} }
\newcommand{\Lammin}{ \Lambda_{\rm min} }
\newcommand{\Lammax}{ \Lambda_{\rm max} }

\title{Constructing Gravitational Dimensions}

\author{Matthew D. Schwartz\\
Jefferson Laboratory of Physics, Harvard
  University, Cambridge, MA 02138 \\ 
email: matthew@schwinger.harvard.edu}

\preprint{HUTP-03/A021\\}

\abstract{
It would be extremely useful to know whether a particular low energy
effective theory might have come from a compactification
of a higher dimensional space. 
Here, this problem is
approached from 
the ground up by considering theories with multiple interacting
massive gravitons.
It is actually
very difficult to construct discrete gravitational dimensions
which have a local continuum limit.
In fact, any model with only nearest neighbor interactions is doomed.
If we could find a non-linear extension for the Fierz-Pauli
Lagrangian for a graviton of mass $m_g$
which does not break down until the scale 
$\Lambda_2=\sqrt{m_g \mpl}$,
this could be used to construct a 
large class of models whose continuum limit is
local in the extra dimension.
But this is shown to be impossible: 
a theory with a single graviton
must break down by $\Lambda_3 = (m_g^2 \mpl)^{1/3}$.
Next, we look at
how the discretization prescribed by the
truncation of the KK tower of an
honest extra diemsinon rasies the scale of strong coupling.
It dictates an intricate set of interactions among various
fields which conspire to soften the strongest
scattering amplitudes and allow for a local continuum limit,
at least at tree-level.
A number of canditate symmetries associated with locality
in the discretized dimension are also discussed.
}
\begin{document}
\begin{fmffile}{sym4pics}

\section{Introduction}
There are many compelling reasons to study discrete gravitational
dimensions. 
The ultimate goal, of course, is to construct
a space-time lattice which reproduces general relativity at
low energies. A more practical application would be towards
phenomenological extensions of the standard model.
Here we use them to
characterize what type of low energy effective theories
might have arisen from the compactification of a continuous extra
dimensional space. 
Until recently, the best approach to this problem seemed
to be to take an extra dimensional model
and work out
the low energy theory by explicitly
integrating out the extra dimensions.
Because the KK tower of such theories can
be truncated at very high energy without harming the low energy
theory, any such model can be interpreted as
a discrete theory space by a simple Fourier transform.
So the question becomes:
which theory spaces produce low energy
effective theories with an extra dimensional
interpretation? 
Because we know how to study such
gravitational theory spaces directly \cite{us1},
we now have a very general approach to the problem. 

Normally, we would expect that
a discrete extra dimension should
look continuous for small enough lattice spacing. 
This is true for gauge theories,
where any haphazardly constructed theory space that
looks continuous at the linear level
can be made to look continuous at the non-linear level if
the discretization is taken sufficiently fine.
That is, violations of unitarity from the gauge boson
interactions can be pushed above the natural cutoff of the
higher dimensional theory by simply shrinking the lattice spacing.
In a recent paper \cite{us2}, it was shown that 
the same simple intuition does not apply for discrete
dimensions involving gravity.
For example, the continuum limit of a discretization
with only nearest neighbor hopping terms must have interactions,
apparent at low energy,
which are highly non-local in the extra dimension.
The problem is that
for gravity self-consistent effective field theory imposes a
limit on how weak we can 
make the unitarity violating effects.
In \cite{us2}, the origin of this impediment was traced to
the crazy scalar longitudinal mode of a massive graviton which
propagates only after mixing with the transverse modes. 
It was also argued in \cite{us2} that the truncated KK theory
of a single compact extra dimensional model {\it does} have a local
continuum limit. However, no explanation was given about how
the scalar longitudinal mode is dealt with from the 
point of view of the low energy effective theory.

In this paper, we begin to explore how to construct theory spaces from the
ground up. We elaborate on the results of \cite{us1} and \cite{us2}
and close the book on a number of issues left unresolved 
by those investigations.
First, we attempt to improve the minimal nearest neighbor
discretization
by adding non-linear, but still nearest neighbor, interactions among the
site gravitons.
If there were a extension of
the Fierz-Pauli Lagrangian for a massive graviton which
controlled the dangerous scalar longitudinal mode, this could
be replicated around the minimal model and the continuum limit
would be drastically improved. In Section \ref{secu1} we
show that no such extension exists. 
In fact, we completely characterize all non-linear
extensions Fierz-Pauli 
and show conclusively that a theory for a single graviton
of mass $m_g$ must break down by $(m_g^2 \mpl)^{1/3}$.
Next, we explore a truncated KK theory for the
case of a circle. We study the interacting Lagrangian of this
theory in great detail and compute all of the strongest tree-level
amplitudes. The most dangerous amplitudes involving the troublesome
scalar longitudinal mode of the lightest massive graviton
are softened by the exchange of heavier gravitons and the massless
graviphoton. The radion also contributes, but not to the strongest
diagrams.
Finally, we make some comments about various broken symmetries
and discuss some implications of this work.
Much of the technical details
are removed to Appendices \ref{appproof} and \ref{appfai};
all of the important qualitative results are presented in the
main text.

\section{Goldstone Bosons and the Minimal Discretization \label{secreview}}
We begin with a review of \cite{us1} and \cite{us2},
with a few added niceties.
The theories we will consider all contain at least one
massive graviton, and therefore involve Lagrangians of
Fierz-Pauli \cite{Fierz:1939ix} form:
\begin{equation}
\cL = 
\mpl^2 \sqrt{g} R ( g ) + 
\mpl^2 m^2_g 
( g_{\mu \nu} - \eta_{\mu \nu} ) 
( \eta^{\mu \rho} \eta^{\nu \sigma} - \eta^{\mu \nu}\eta^{\rho \sigma} ) 
( g_{\rho \sigma} - \eta_{\rho \sigma} )
+ \cdots
\label{lmin}
\end{equation}
The mass term explicitly breaks general coordinate invariance (GC)
and leads to the propagation of the longitudinal modes of the
graviton. 
It is helpful to project out these modes directly
as separate fields which can be interpreted as the Goldstone
bosons for the breaking of the GC symmetry. 
This is done by applying the coordinate transformation
$x^\alpha\to y^\alpha(x)=x^\alpha+\pi^\alpha(x)$ 
to the Lagrangian. The
dependence of the new Lagrangian on the Goldstone bosons $\pi^\alpha$
conveys all the effects of the broken symmetry. More explicitly,
we apply the following replacement to \eqref{lmin}:
\begin{equation}
g_{\mu \nu} \rightarrow \tilde{g}_{\mu \nu} \equiv \frac{\partial
y^{\alpha}}{\partial x^{\mu}} \frac{\partial y^{\beta}}{\partial x^{\nu}}
g_{\alpha \beta} ( y ) = ( x^{\alpha} + \pi^{\alpha} )_{, \mu} ( x^{\alpha}
+ \pi^{\beta} )_{, \nu} g_{\alpha \beta} ( x + \pi ) \label{piints}
\end{equation}
It also follows that the Lagrangian which results from this
replacement is generally coordinate invariant. After all, 
the $\pi^\alpha$ represent {\it all} the symmetry violating
effects. Of course, $\pi^\alpha$ must transform non-linearly,
but its transformation law is simply induced from the transformation
of $y^\alpha$ and given in \cite{us2}.

%It is important to observe that this way of introducing Goldstone bosons is
%not unique. We could equally well have used
%\begin{equation}
%\cL = \mpl^2 \sqrt{g} R ( g ) + 
%\sqrt{g} ( g_{\mu \nu} - \eta_{\mu \nu} ) ( g^{\mu \rho} g^{\nu
%\sigma} - g^{\mu \nu} g^{\rho \sigma} ) ( g_{\rho \sigma} - \eta_{\rho
%\sigma} ) \label{lmax}
%\end{equation} 
%and replaced
%\begin{equation}
%\eta_{\mu \nu} \rightarrow \tilde{\eta}_{\mu \nu} \equiv \frac{\partial
%y^{\alpha}}{\partial x^{\mu}} \frac{\partial y^{\beta}}{\partial x^{\nu}}
%\eta_{\alpha \beta}
%\end{equation} 
%Now the $y^{\alpha}$ have to transform so that $\tilde{\eta}_{\mu \nu}$ is
%covariant. No matter which way we introduce the Goldstine bosons, the 
%physics must be the same 
%if it has the same Lagrangian in the unitary gauge. Note
%that \eqref{lmin} and \eqref{lmax} are not the same Lagrangian. \eqref{lmax}
%has additional
%interactions. 

At this point, it is useful to expand the metrics around flat space
$g_{\mu \nu} = \eta_{\mu\nu} + h_{\mu \nu}$ and
the Goldstone bosons as $\pi^\alpha = \eta^{\alpha\mu} A_\mu$. 
Then, after an integration by pars, 
the Lagrangian \eqref{lmin} becomes:
\begin{equation}
\cL =
\frac{1}{4}\mpl^2( -h_{\mu \nu , \alpha}^2 
+ 2 h_{\mu\nu , \mu}^2 
- 2 h_{\mu \nu , \mu} h_{,\nu} 
+ h_{,\mu}^2)
+ 
\mpl^2 m_g^2
(F_{\mu \nu}^2 + 2
A_{\mu , \nu} h_{\mu \nu} - 2 A_{\mu , \mu} h ) + \cdots \label{hkin}
\end{equation}
The first part is the standard kinetic term for a massless
graviton, 
and the second part contains the standard kinetic term
for $A_\mu$. Note that if we had chosen a tensor structure
for the mass term in \eqref{lmin} different from Fierz-Pauli,
$A_\mu$ would have non-standard kinetic term 
signalling unitary violation at $\sim m_g$.

$A_\mu$ is an interacting vector boson, which for lack of any gauge
symmetry, contains three propagating degrees of freedom.
We can separate out its longitudinal mode, which corresponds
to the scalar longitudinal mode of $\hmn$ by substituting 
$A_{\mu}\rightarrow A_{\mu} + \phi_{,\mu}$.
This establishes an
artificial $U(1)$ symmetry
for which $\phi$ is the Goldstone boson. We will return to this 
symmetry in Section \ref{secu1}.
Using a more schematic notation, the Lagrangian becomes:
\be
\cL =
 \mpl^2 h \square h + 
\mpl^2 m^2_g ( A \square A + h \square \phi ) +
\cdots 
\ee
In this Lagrangian, $\phi$ only gets a kinetic term from mixing
with $h$. Naturally, because $\phi$ always appears with two derivatives,
the only way it could get a proper kinetic term is through mixing.
Nevertheless, this feature is
the source of all the bizarre features
of massive gravitons discussed in \cite{us1,us2} and expounded here.

To study the interacting theory, we need to canonically normalize
the fields:
$h_{\mu \nu}^c = \mpl h_{\mu \nu}$,
$A_{\mu}^c = m_g \mpl A_{\mu}$, and
$\phi^c = m_g^2 \mpl \phi$.
Thus, each interaction will have an associated scale which 
we can read directly off the Lagrangian. Because all the strong
interactions, involving the Goldstone fields,
come out of the mass term in \eqref{lmin}, we 
can derive a general formula:
\begin{equation}
m_g^2 \mpl^2 A_{\mu}^{n_A} \phi^{n_{\phi}} h^{n_h} 
= \Lambda_\lambda^{4 - n_A - n_{\phi} 
- h_h} A^{c n_A}_{\mu} \phi^{c n_{\phi}} h^{c
n_h}
\end{equation} 
where
\begin{equation}
\Lambda_{\lambda} = \left(m_g^{\lambda-1} \mpl \right)^{1/\lambda}, 
\quad\quad
\lambda = \frac{3 n_{\phi} + 2 n_A + n_h - 4}
{n_{\phi} + n_A + n_h - 2} 
\label{lamform}
\end{equation}
This implies, for example, that the strongest vertex 
is $\phi^3$ which has the scale
$\Lambda_5= (m_g^4 \mpl)^{1/5}$. The amplitude for
a  simple exchange diagram
involving this vertex will grow as $\cA \sim E^{10}/\Lambda_5^{10}$.

Incidentally, it may seem strange that the Lagrangian \eqref{lmin}
should be based on $\sqrt{g}R$ when general coordinate invariance is
explicitly broken by the mass term.
But this partial GC symmetry guarantees that all the interactions
coming from the $\sqrt{g}R$ term  involve only transverse
polarizations.
If this were not true, and a term like $\mpl^2 \partial^2 h^3$ were present
with arbitrary tensor structure, it would produce interactions of $\phi$
which a simple calculation shows
get strong at $\Lambda_7$. 
So, the GC symmetry in the $\sqrt{g}R$ term,
which has {\it all} the interactions in unitary gauge,
actually raises the scale of strong coupling.
While this is not a qualitative improvement, it does
demonstrate that $\Lambda_5$ is not the lowest possible
scale where a theory for single massive graviton based on Fierz-Pauli
could break down.
In fact, the whole point of introducing Goldstone bosons as
a symmetry breaking effect is that 
we can start at $\Lambda_5$; the cancellation of the $\Lambda_7$ diagrams,
which would be obscure in unitary gauge, is given for free.

We continue our review by looking at the minimal lattice 
explored in \cite{us2}. The theory space picture looks like:
\be
%
% 5 sites space
%
\parbox{30mm}{
\begin{fmfgraph*}(70,70)
\fmfsurroundn{v}{5}\fmfdotn{v}{5}
\fmfv{decoration.shape=circle,
decoration.filled=empty,decoration.size=0.2w}{v1}
\fmfv{decoration.shape=circle,decoration.filled=empty,
decoration.size=0.2w}{v2}
\fmfv{decoration.shape=circle,decoration.filled=empty,
decoration.size=0.2w}{v3}
\fmfv{decoration.shape=circle,decoration.filled=empty,
decoration.size=0.2w}{v4}
\fmfv{decoration.shape=circle,decoration.filled=empty,
decoration.size=0.2w}{v5}
\fmf{fermion}{v1,v2}
\fmf{fermion}{v2,v3}
\fmf{fermion}{v3,v4}
\fmf{fermion}{v4,v5}
\fmf{dots}{v5,v1}
\end{fmfgraph*} }
\ee
The associated Lagrangian is simply \eqref{lmin} with
the mass terms replaced by hopping terms:
\be
\Lmin =\sum_j M^2 \sqrt{g^j}R[g^j] + M^2 m^2 \sqrt{g^j} 
( g^j_{\mu \nu} - g^{j + 1}_{\mu \nu} ) 
( g_j^{\mu \rho} g_j^{\nu \sigma} - g_j^{\mu \nu} g_j^{\rho \sigma} ) 
( g^j_{\rho \sigma} - g^{j + 1}_{\rho \sigma} ) \label{lione}
\ee
The hopping terms break all but one of the 
general coordinate invariances. So we restore these symmetries by
by replacing:
\begin{equation}
g^{j+1}_{\mu \nu} \rightarrow 
\frac{\partial y_j^{\alpha}}{\partial x^{\mu}} 
\frac{\partial y_j^{\beta}}{\partial x^{\nu}} 
g^{j+1}_{\alpha \beta} ( y_j ) \label{gjsub}
\end{equation}

Next, we expand metrics around flat space and the $y_j$ in terms
of vector and scalar Goldstones $a_\mu^j$ and $\phi^j$ (using lower case and
$j$ for the site basis).
Then the Lagrangian looks like:
\begin{equation}
\Lmin =   M^2 h_j \square h_j + M^2 m^2
\{ (h_j-h_{j+1})^2 +  (h_j-h_{j+1}) \square \phi_j +
a_j \square a_j + \phi_j\phi_j\phi_j+ \phi_j a_j a_j\} + \cdots 
\label{lminsite}
\end{equation}

To diagonalize the mass matrix, we take the standard linear combinations:
$h_j = e^{ 2\pi i \frac{j n}{N} } G_n$, 
$a_j = e^{ 2\pi i \frac{j n}{N} } A_n$, and
$\phi_j = e^{ 2\pi i \frac{j n}{N} } \Phi_n$ 
(uppercase and $n$ for the momentum basis).
Then, summing over $j$, and using the approximation 
$m_n \sim m \frac{n}{N}$
the Lagrangian becomes:
\begin{multline}
\Lmin = N M^2 G_n \square G_{-n} \\
+ N M^2 m^2
\{\frac{n^2}{N^2} G_n G_{-n} + \frac{n}{N} G_n \square \Phi_{-n} +
A_n \square A_{-n} + \Phi_n \Phi_m \Phi_{-n-m} +\Phi_n A_m A_{-n-m} +
\cdots \} \nonumber
\end{multline}
Just as with a single massive graviton, we can read of the strength
of the interactions after going to canonical normalization:
$G_n = \frac{1} {\sqrt{N} M} G_n^c$,
$A_n = \frac{1}{\sqrt{N} M m } A_n^c$,
and $\Phi_n = \frac{\sqrt{N}}{n M m^2}\Phi^c_n$.
In terms of the 
the physical scales $\mpl=M\sqrt{N}$ and $m_1 = m/N$
the strongest interactions look like:
\begin{equation}
\cL = \cdots + \frac{1}{N \mpl m_1^4} \Phi_1^c \Phi_1^c \Phi_{-2}^c+
\frac{1}{N \mpl m_1^2} \Phi^c_1 A^c_1 A^c_{-2} + \cdots \label{cnint}
\end{equation}
We then read off that the strong coupling scale, set
by the $\Phi^3$ vertex is:
\be
\Lammin = (N m_1^4 \mpl)^{1/5} \label{lammin}
\ee

This scale seems reasonable. 
Formally, $\Lammin$ goes to $\infty$ as $N\to\infty$, and
so we can reproduce linearized 5D gravity at low energy. However,
within a consistent effective field theory, we
can never take $\Lammin$ higher than the mass of the heaviest
modes in the theory $m_N \sim N m_1$. This constraint, can
be written as:
\be
\Lammin < \Lammax = \mv (R \mv )^{-5/8} \label{minmax}
\ee
where $R=1/m_1$ is the size of the discrete dimension and
$\mv=(m_1 \mpl^2)^{1/3}$ is the 5D Planck scale. Since 
$\Lammax$ must be less than $\mv$ this theory has no hope
of looking like 5D gravity in the continuum limit.

Nevertheless, there is nothing wrong with taking $N \to \infty$
keeping $\mpl$ fixed in the minimal discretization. 
The resulting continuum theory will
be a consistent effective field theory, even if it
cannot be interpreted as having a smooth extra dimension.
The argument in \cite{us2} for
why the continuum limit will be non-local can be paraphrased as
follows.
The interactions in  \eqref{lminsite}
are in terms of $\phi_j$, but $\phi_j$
gets a kinetic terms from coupling to 
$h_j - h_{j+1} = \Delta_z h_j$. 
Equivalently, $\Phi=\Delta_z \phi_j = \phi_j - \phi_{j+1}$
is the physical, propagating
field. So the dangerous interactions
are really $\phi^3_j \sim \frac{1}{\Delta_z^3}\Phi^3$ which have a  non-local
continuum limit.

\section{Improving the Minimal Model \label{secu1}}
The simplest improvement on the minimal discretization
would be a model which still only has nearest neighbor interactions,
but whose unitary gauge Lagrangian 
is a more complicated function of $g^j_{\mu\nu} - g^{j+1}_{\mu\nu}$.
These models are particularly easy to study, because all of their
features can be understood from simply looking at non-linear extensions
of the Fierz-Pauli Lagrangian for a single massive graviton.
Of course, it is unlikely that this approach will provide
a significant improvement over the minimal discretization,
because these modifications are still strictly local,
and we are trying to cure a non-local disease.
Nevertheless, this fairly clean set of models will
help us understand the locality problem within the low
energy field theory. And if they were to succeed (which they won't)
we would have all the freedom to construct gravitational theory
spaces that we have for gauge theory spaces.

To begin, we should address the question of what property
of the effective theory guarantees locality in the continuum limit.
Recall that the obstacle to taking a smooth continuum limit 
of the minimal
discretization is that all the modes are not necessarily
lighter than the cutoff, a {\it sine qua non} of a consistent
effective field theory. 
No such restriction exists for
a weakly coupled gauge theory because the cutoff
$\Lambda \sim 4\pi N m_1/ g$ is {\it always} above the top of
the tower: $\Lambda > m_N \sim N m_1$. 
For gauge theory, the guarantee follows from the fact that $\Lambda$ depends on
$N$ and $m_1$ only through the product $N m_1 \sim m_N$.
In contrast, for gravity
the scale  $\Lammin = (N m_1^4 \mpl)^{1/5}$ does not depend
on $N$ and $m_1$ in an auspicious combination. 
%In guage theory,
%when we take the Lagrangian for a single gauge boson of mass $m_A$
%with cutoff $\Lambda = 4\pi m_A/g$ and replicate it around the theory
%space, we get a factor of $N$
%with 
However, if we could find a Lagrangian
for a single massive
graviton which breaks down at 
$\Lambda = \Lambda_2 = \sqrt{ m_g \mpl }$
then the discretization based on this Lagrangian would
have $\Lambda = \sqrt{N m_1 \mpl}$ and would automatically satisfy
the consistency constraint. This theory would have a local
continuum limit.
So our task becomes simply: extend the Lagrangian for a massive
graviton so that it gets strong at $\Lambda_2$ (or higher).

Note, in passing, that the $\Lambda_2$ scale for a single
massive graviton is the geometric mean
between $\mpl$ and $m_g$. 
In particular, if we take the graviton to have
a Hubble mass $m_g \sim H$, then $\Lambda_2 \sim $mm$^{-1}$,
which happens to be the current limit to which gravity
has been experimentally probed. Of course, the graviton could
not have a Hubble mass because of other constraints from non-linear
effects around large massive sources, as discussed at length in \cite{us1}.
But if we look only at short distance constraints, raising the scale
for strong coupling of 
a single massive graviton to $\Lambda_2$ would be absolutely necessary
to avoid obvious
 contradiction with experiment.

Now, any Lagrangian we consider must start with a quadratic
term of Fierz-Pauli form:
\be
\LFP = h_{\mu\nu}^2 - h^2 \label{lfpu}
\ee
Since we already understand kinetic mixing, and
are not presently interested in the relatively
weakly coupled transverse modes, let us introduce
the Goldstones as in \eqref{piints} and then set
$\hmn=0$. We will be making this transformation
often for the rest of the paper and denote it
$\leadsto$. It is 
equivalent to replacing:
\be
\hmn \leadsto A_{\mu,\nu} + A_{\nu,\mu} + A_{\alpha,\mu}A_{\alpha,\nu}
\ee
Thus, after an integration by parts
\be
\LFP \leadsto -F_{\mu\nu}^2 - 4 A_{\mu,\mu} A_{\mu,\nu} A_{\mu,\nu} 
+ 4 A_{\mu\nu}A_{\nu\alpha}A_{\mu\alpha}
+ \cdots
\label{lfpt}
\ee

The appearance of $F_{\mu\nu}^2$ in \eqref{lfpt} is suggestive.
Recall that $\phi$, which we have not yet introduced into \eqref{lfpt},
is the longitudinal mode of the vector field $A_\mu$. It is
the Goldstone boson for the breaking of a ``fake'' $U(1)$ symmetry which
\eqref{lfpt} has already at quadratic level. Of course, we cannot
expect the entire Lagrangian to have a $U(1)$ symmetry, because
$\phi$ is necessary to reproduce 5D gravity at the linear level.
But we might hope that by
adding cubic and higher order terms to \eqref{lfpu}, we can
achieve a gauge invariance in the Goldstone 
Lagrangian, that is, the Lagrangian after the $\leadsto$
transformation.
In other words, 
the Fierz-Pauli structure may be the first part of an expansion
fixed by $U(1)$ gauge invariance of the vector longitudinal modes.
Moreover, 
we can see from \eqref{lamform} that all the interactions we are
trying to get rid of, the ones
which get strong below $\Lambda_2$, involve the field $\phi$. So
this symmetry condition is sufficient for the construction of
discretizations with local continuum limits.

Alas, it turns out that the $U(1)$ is a complete red-herring.
We will now see  not only that the $U(1)$ invariance
embedded in the Fierz-Pauli structure is restricted to the quadratic terms, 
but, more strongly, that there is no way to raise the scale of 
strong coupling for
a single massive graviton higher than $\Lambda_3$.

\subsection{Extending Fierz-Pauli}
At this point, it is handy to introduce some notation.
The vector of Goldstones, $A_\mu$, will always come with a
derivative, so we can represent $A_{\mu,\nu}$ as a matrix:
\be
A &\equiv& A_{\mu , \nu} \Rightarrow A^T = A_{\nu , \mu} \label{amatnot}\\
F &\equiv& A_{\mu , \nu} - A_{\nu , \mu}
= A - A^T \\
\Phi &\equiv& \phi_{, \mu , \nu} = \Phi^T \label{phimat}\\ 
1 &\equiv& \eta_{\mu\nu}
\ee
Projecting out the longitudinal modes from a given unitary gauge
Lagrangian amounts to replacing:
\begin{equation}
h_{\mu\nu} \leadsto 
%\eta_{\mu \nu} - \frac{\partial y^{\alpha}}{\partial x^{\mu}} \frac{\partial
%y^{\beta}}{\partial x^{\nu}} \eta_{\alpha \beta} = 1 - ( 1 + A^T ) ( 1 + A )
A + A^T + A^T A \label{apa}
\end{equation} 
%Any Lagrangian which can be written as an expansion in $h$ in unitary
%gauge will depend only on this combination of Goldstones 
%after the transverse modes are set to zero.
Also, since, by Lorentz invariance, we will
always be taking traces of such matrices, we define the $[\cdots]$
notation by:
\begin{equation}
[A \cdots A] \equiv 
\mathrm{Tr}\; [A \cdots A] = A_{\mu, \nu} \cdots A_{\alpha,\mu}
\end{equation} 
So, in the new notation, the Fierz-Pauli term becomes:
\be
\LFP &=& [h^2] - [h]^2 \nn \\
&\leadsto&
[( A + A^T + A^T A )^2] - ( 2 [A] + [A A^T])^2 \nn \\
&=& 2[A A^T]+2[A A^T]-4[A]^2 + 4 [A^2 A^T] - 4 [A] [A A^T] + [A^T A A^T A] - 
[A A^T]^2 \quad \quad \label{l2int}
\ee
The third line involved an integration by parts.
We can express this in terms of the symmetric and antisymmetric parts
of $A$: $\Phi$ and $F$. 
That is, we set $A=\Phi+F$ and $A^T = \Phi-F$. 
Then,
\begin{equation}
\LFP \leadsto 
-4[F^2] 
+4 [\Phi^3] - 4 [\Phi] [\Phi^2]
+[F^4] 
-[F^2]^2
+ [\Phi^4]
-[\Phi^2]^2
- 4[\Phi^2 F^2] + 2[\Phi F\Phi F]
+2[F^2][\Phi^2]\nn
\end{equation}

With this notation, it will be much easier to study extensions
of Fierz-Pauli.
The $U(1)$ symmetry we are searching for implies
that the Goldstone Lagrangian
should depend only on $F$ and not on $\Phi$.
First, observe that 
\begin{equation}
h \leadsto 2 \Phi + \Phi^2+\Phi F - F \Phi - F^2
\end{equation}
Since the only first order term in the expansion of $h$ is $\Phi$,
we can always eliminate the $\Phi$ self-couplings
from the Lagrangian. For example, we cancel the third order terms by adding 
\begin{equation}
\cL_3 = - \frac{1}{2} [h^3] + \frac{1}{2} [h] [h^2]
\end{equation} 
Thus the lowest order $\Phi$ self couplings in 
$\cL_{FP} + \cL_3$ will be $\Phi^4$. These can be
eliminated by adding  an $\cL_4$ with quartic terms, and so on.
By induction, it is easy to see that all the self-interactions of the scalar
can be eliminated in this way.

The next order gauge-violating terms look like $F^2\Phi^2$. 
Up to fourth order, there are 12 terms we must eliminate:
\begin{equation}
[\Phi^3], [\Phi]^2[\Phi], [\Phi]^3, [\Phi^4],
[\Phi^3][\Phi], [\Phi^2]^2, [\Phi^2][\Phi]^2, [\Phi]^4,
[F^2\Phi^2],[F\Phi F\Phi], [F^2][\Phi^2], [F^2][\Phi]^2\label{allpf}
\end{equation}
These are related by two equations that come from integration
by parts:
\begin{equation}
[\Phi^3] = 2[\Phi]^2+[\Phi]^3 \quad\text{and}\quad
[\Phi^4] = [\Phi]^3[\Phi] +[\Phi^2][\Phi^2] + [\Phi^2][\Phi]^2+ [\Phi]^4
\end{equation}
So there are 10 independent terms which must vanish. However,
the most general Lagrangian up to fourth order in $h$ has only
8 terms:
\begin{equation}
\cL_{\Delta} 
= c_1 [h^3] + c_2 [h^2] [h] + c_3 [h]^3 + q_1 [h^4] + q_2 [h^2] [h^2]
+ q_3 [h^3] [h] + q_4 [h^2] [h]^2  + q_5 [h]^4 \label{hcq}
\end{equation}
We might also consider terms with derivatives acting on $h$,
but these cannot produce terms of the form  \eqref{allpf}.
Therefore, unless there is some special arrangement,
we do not have enough freedom to fabricate a $U(1)$ symmetry.

Still, it may be possible that although the Lagrangian is
gauge dependent, 
all the physical scattering
processes involving the $\phi$ fields vanish. 
This could
be understood as a non-linearly realized $U(1)$ symmetry
which is obscured by our choice of the transverse and
longitudinal modes of $A_\mu$. For example, 
the field redefinition $A_\mu \to B_\mu+B_\beta B_{\mu,\beta}$
will produce interactions in the non-interacting Lagrangian
$(A_{\mu,\nu}-A_{\nu,\mu})^2$. 
These interactions will not vanish by integration
by parts, but all the physical 
scattering amplitudes involving the new $B_\mu$ fields will be zero.
We could certainly try to classify all non-linear field redefinitions,
and all other reasons that the amplitudes may vanish while the
interactions do not. But it is more straightforward just to compute
the dependence of the strongest scattering amplitudes
on the coefficients in \eqref{hcq}.
This is done in Appendix \ref{appproof}.
The conclusion is that
it is simply impossible to extend the Fierz-Pauli Lagrangian so
that unitarity is preserved for a single massive graviton 
above $\Lambda_3$.

\section{Truncated KK Theory \label{secKK}}
We have seen that it is impossible to eliminate all the dangerous
amplitudes for scattering of the scalar longitudinal modes of
a massless graviton by a non-linear extension of the Fierz-Pauli
Lagrangian. Had this been possible, we could have used
the extended Lagrangian as a template for the link structure
in a discretization with only nearest neighbor interactions,
and the resulting theory would have had none of the problems
of the minimal model discussed in \cite{us2}.
This failure is disappointing, but could have been anticipated
from the fact that the continuum sickness of the minimal model
is non-locality in the extra dimension, so its cure should
involve non-locality on the lattice.

Despite the discouraging results of Section \ref{secu1}, 
we know that the scale
of strong coupling for a discrete extra dimension can be raised
above $\Lammin = (N m_1^4 \mpl)^{1/5}$.
As was mentioned in \cite{us2}, a truncated KK theory {\it does}
have a local continuum limit, in contrast to the minimal discretization.
This simply follows from the observation that a truncation performed
at high energy cannot effect low energy physics without undermining
the general assumptions of effective field theory.
Somehow the low KK modes are not interacting strongly at $\Lambda_5$
or even at $\Lambda_3$.

In this section, we study the
truncated KK theory in great detail. We work out the 
Lagrangian in theory space, including the radion, graviphoton,
and all the tensor structure. Its non-locality, including the power-law
decay of the interactions with distance in the discrete dimension,
is apparent. There are a number of subtle issues about the introduction
of Goldstone bosons which are also explored. Finally, the
amplitudes for all
the dangerous diagrams involving the scalar longitudinal polarizations
of the lowest KK modes are given. Due to exchange of heavier modes, and,
somewhat surprisingly, the radion and graviphoton as well, all the
dangerous amplitudes cancel at tree level.

Start with a 5D metric $G_{M N}$. We will label the compact fifth
direction as $z$ and the non-compact directions collectively as $x$. Then we
gauge fix as much as possible, so the metric takes the form
\begin{equation}
g_{5 D} = \left( \begin{array}{cc} g_{\mu \nu} ( x , z ) + e^{2 r ( x )}
V_{\mu} ( x ) V_{\nu} ( x ) & e^{2 r ( x )} V_{\mu} ( x )\\ e^{2 r ( x )}
V_{\nu} ( x ) & e^{2 r ( x )}\\ \end{array} \right)  \label{l5d4d}
\end{equation}
$r$ is the radion and $V_\mu$ is the graviphoton. In this gauge, neither
$r$ nor $V_\mu$ depends on $z$ and
the Lagrangian becomes:
\be
\cL
&=& \mv^3 \sqrt{g_{5 D}} R_{5 D} ( g_{5 D} )\\ 
&=& \mv^3 \sqrt{g ( x , z )} 
\Big{\{} 
e^r R_{4 D} ( g ) + \frac{1}{4} e^{- r} (
\partial_z g_{\mu \nu} ( g^{\mu \rho} g^{\nu \sigma} - g^{\mu \nu}
g^{\rho \sigma} ) \partial_z g_{\rho \sigma} ) \label{lkkmid}\\
&&\quad\quad\quad\quad\quad
\hskip 1cm
- \frac{1}{4} ( V_{\mu , \nu} - V_{\nu , \mu} ) g^{\mu \nu} g^{\rho
\sigma} ( V_{\rho , \sigma} - V_{\sigma , \rho} )
+  \cL_V 
\Big{\}}
\label{lvintro}
\ee 
Here 
$R_{5 D}(g_{5 D})$ is the
five dimensional Ricci scalar constructed from 5D metric. 
$R_{4 D}(g)$ is the 4D
Ricci scalar constructed out of $g_{\mu \nu} ( x , z )$ which
we treat as a 4D metric which happens to depend on a continuous parameter
$z$. $\cL_V$ contains interactions of the graviphoton with the other
fields, some of which are presented in
Appendix \ref{appfai}.

Now, we assume the compact dimension is a circle
and expand the metric in KK modes: 
$g_{\mu \nu} ( x , z ) = 
\sum G_{\mu\nu}^j ( x ) e^{2\pi i \frac{j z}{R}}$.
We continue the convention that lower-case fields are in the site basis
and uppercase fields are in the KK basis. Then $\cL$ contains
the mass terms:
\begin{equation}
\cL =
\mpl^2 m_1^2 n^2 ([G_n G_{-n}] - [G_n] [G_{-n}] )+ \cdots \label{l2gg}
\end{equation}
The scale for the masses is set by the radius:
$m_1 = 2 \pi/R$; and we have introduced the effective 4D Planck scale
of the low energy theory: $\mpl^2 = \mv^3 R$. 
We see that the spectrum comprises a massless graviton,
a doubly degenerate tower of massive gravitons, and
the massless radion and 
graviphoton.\footnote{It might seem that the theory would have been
simpler if we had compactified on an interval instead of a circle,
thereby removing the mass degeneracy and the graviphoton. In fact,
because the KK wavefunctions for the interval are sines instead
of exponentials, the interactions in the circle are much easier
to work with. This point is discussed further in Appendix \ref{appfai}.
}
\FIGURE[t]
{\centerline{\epsfig{file=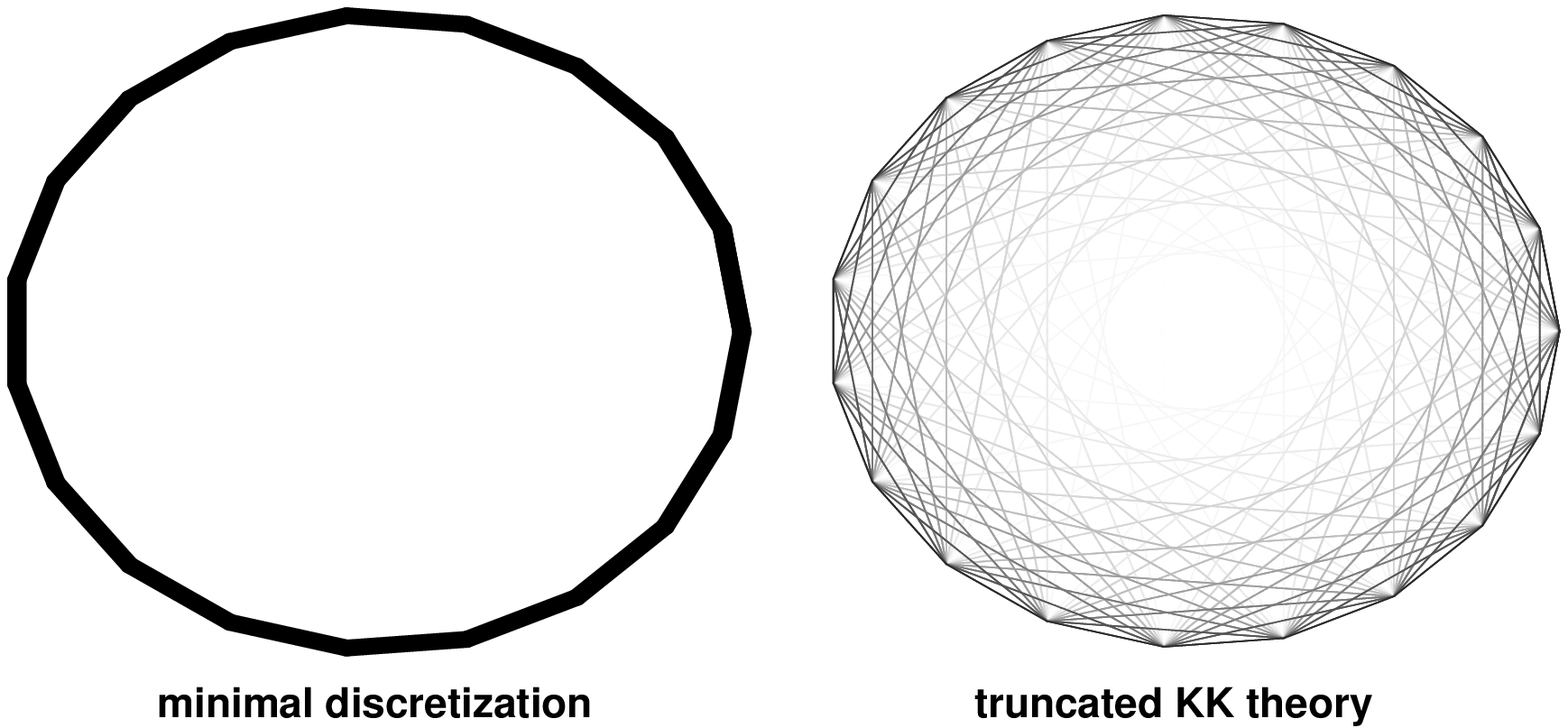,width=\textwidth}}%
\caption{In the minimal deconstruction links are only between
nearest neighbors. In the site basis for the truncated KK theory,
there are
links between every pair of sites, but the strength of the link
dies off with distance. These links which are non-local in theory
space remain in the limit of a large number of sites.
\label{mdfdfig} }
}

Now we truncate the theory to $N$ modes,
and go back to position space via
the discrete Fourier transform
$G_n = \frac{1}{N} e^{2 \pi i \frac{n a}{N}} g_a$.
Then, the mass terms become:
\begin{equation}
\cL_2 = n^2 G_n G_{- n} = \frac{n^2}{N^2} e^{2 \pi i n
\frac{\Delta a}{N}} g_a g_{a + \Delta a} = \frac{2 n^2}{N^2} \cos
\frac{2 \pi n \Delta a}{N} g_a g_{a + \Delta a}
\end{equation} 
Evidently, there are links between distant sites. 
As $N \rightarrow \infty$
\begin{equation}
\cL_2 \rightarrow \frac{1}{6 \pi^2} N ( 2 g_a^2 - \frac{1}{\Delta a^2} 
g_a g_{a + \Delta a} )
\end{equation} 
Even in this limit, 
the interactions which are non-local in theory space are only
power-law suppressed.
This is contrasted to the case of the minimal discretization in
Figure \ref{mdfdfig}.

Now let us look at the interactions.
We know that in a continuum
limit, the theory would break down at the 5D Planck scale
$\mv=\Lambda_{3/2}=(m_1 \mpl^2)^{1/3}$. 
We also know that if we truncate to two modes, a
massless and a massive one, the theory would break down {\it at least} by 
$\Lambda_3=(m_1^2 \mpl )^{1/3}\ll\Lambda_{3/2}$. This is due to scattering
of the first massive mode, with mass $m_1$. So the additional modes
must somehow cancel the strongest diagrams involving mode $1$. 
Moreover,
we know which fields may contribute to this cancellation. Momentum conservation
in the fifth direction translates to KK number conservation in the 4D
theory. So the only fields which can contribute to tree-level
scattering of the
gravitons at the first mass level are gravitons at the second mass level
(with mass $m_2=2m_1$),
the massless graviton, the graviphoton, and the radion. Therefore, we
are interested in the tree level processes:
\begin{equation}
%
% 1111 blob
%
\parbox{25mm}{
\begin{fmfgraph*}(70,40)
\fmfset{curly_len}{2.5mm}
\fmfleft{p1,p2}
\fmfright{q1,q2}
\fmf{gluon}{p1,p}
\fmf{gluon}{p2,p}
\fmf{gluon}{p,q1}
\fmf{gluon}{p,q2}
\fmfblob{0.3w}{p}
\fmfv{l=$1$,l.a=120,l.d=.05w}{p1}
\fmfv{l=$1$,l.a=-120,l.d=.05w}{p2}
\fmfv{l=$1$,l.a=60,l.d=.05w}{q1}
\fmfv{l=$1$,l.a=-60,l.d=.05w}{q2}
\end{fmfgraph*} } =
%
%  1111 quartic 
%
\parbox{25mm}{
\begin{fmfgraph*}(60,40)
\fmfset{curly_len}{2.5mm}
\fmfleft{p1,p2}
\fmfright{q1,q2}
\fmf{gluon}{p1,p}
\fmf{gluon}{p2,p}
\fmf{gluon}{p,q1}
\fmf{gluon}{p,q2}
\fmfv{l=$1$,l.a=120,l.d=.05w}{p1}
\fmfv{l=$1$,l.a=-120,l.d=.05w}{p2}
\fmfv{l=$1$,l.a=60,l.d=.05w}{q1}
\fmfv{l=$1$,l.a=-60,l.d=.05w}{q2}
\end{fmfgraph*} }  +
%
%  2-exchange
%
\parbox{25mm}{
\begin{fmfgraph*}(60,40)
\fmfset{curly_len}{2.5mm}
\fmfleft{p1,p2}
\fmfright{q1,q2}
\fmf{gluon}{p1,pl}
\fmf{gluon}{p2,pl}
\fmf{gluon,label=$0,,2$}{pl,pr}
\fmf{gluon}{pr,q1}
\fmf{gluon}{pr,q2}
\fmfv{l=$1$,l.a=120,l.d=.05w}{p1}
\fmfv{l=$1$,l.a=-120,l.d=.05w}{p2}
\fmfv{l=$1$,l.a=60,l.d=.05w}{q1}
\fmfv{l=$1$,l.a=-60,l.d=.05w}{q2}
\end{fmfgraph*} }  +
%
%  graviphoton -exchange
%
\parbox{25mm}{
\begin{fmfgraph*}(60,40)
\fmfset{curly_len}{2.5mm}
\fmfleft{p1,p2}
\fmfright{q1,q2}
\fmf{gluon}{p1,pl}
\fmf{gluon}{p2,pl}
\fmf{photon,label=$V$}{pl,pr}
\fmf{gluon}{pr,q1}
\fmf{gluon}{pr,q2}
\fmfv{l=$1$,l.a=120,l.d=.05w}{p1}
\fmfv{l=$1$,l.a=-120,l.d=.05w}{p2}
\fmfv{l=$1$,l.a=60,l.d=.05w}{q1}
\fmfv{l=$1$,l.a=-60,l.d=.05w}{q2}
\end{fmfgraph*} } +
%
%   radion exchange
%
\parbox{25mm}{
\begin{fmfgraph*}(60,40)
\fmfset{curly_len}{2.5mm}
\fmfleft{p1,p2}
\fmfright{q1,q2}
\fmf{gluon}{p1,pl}
\fmf{gluon}{p2,pl}
\fmf{plain,label=$r$}{pl,pr}
\fmf{gluon}{pr,q1}
\fmf{gluon}{pr,q2}
\fmfv{l=$1$,l.a=120,l.d=.05w}{p1}
\fmfv{l=$1$,l.a=-120,l.d=.05w}{p2}
\fmfv{l=$1$,l.a=60,l.d=.05w}{q1}
\fmfv{l=$1$,l.a=-60,l.d=.05w}{q2}
\end{fmfgraph*} }  \label{allex}
\end{equation}

The Goldstone boson formalism lets us analyze
the diagrams in \eqref{allex} very efficiently. We
can classify the amplitudes according to their energy dependence,
{\it i.e.} whether they get strong at $\Lambda_5$, $\Lambda_4$, $\Lambda_3$, 
{\it etc.}. Because of the result from Section \ref{secu1}, that a single
massive graviton must break down by $\Lambda_3$, it is illuminating
to see how the $\Lambda_3$ and stronger diagrams
in the truncated KK theory cancel.
This is demonstrated in explicit detail in  Appendix \ref{appfai}.
As a brief summary, 
we find that
there are no vertices that contribute to $\Lambda_5$;
the quartic vertices which contribute to $\Lambda_4$ are
canceled by exchange of the graviphoton; And both the graviphoton and
the vector polarization of mode $2$
are necessary to cancel the
$\Lambda_3$ diagrams involving external vectors.
This is very important, because it
implies that we are very unlikely to find a compactification in which
the second lightest massive KK mode is parametrically heavier than
the lightest one. It also implies that the first tree level diagrams
which the truncation affects involve the $\Lambda_3$ scale
appropriate external modes at mass level $N/2$.
That is, the truncated KK theory gets strong at 
$\Lambda \sim (N^2 m_1^2 \mpl)^{1/3}$.
It is worth pointing out that
the calculations in the appendix are a  
highly non-trivial check on the goldstone boson equivalence
theorem for gravity and the entire effective field theory for massive
gravitons.

\section{Discussion and Outlook}
In this paper, a number of important
features of theories with massive gravitons have been analyzed. 
An arbitrary theory space which has only nearest neighbor
interactions will lead, in the continuum limit, 
to a theory which looks extra-dimensional at
the linear level, but has highly non-local interactions.
This was understood to be a result of the anomalously strong
scattering amplitudes for the light gravitons in the theory.
We attempted to eliminate these amplitudes with a non-linear
extension of the Fierz-Pauli Lagrangian, but found that
no such extension could push the strong coupling scale
above $\Lambda_3$. We would need to push it as high as $\Lambda_2$
to improve the continuum limit of nearest neighbor discretizations.
In contrast, a theory space based on the truncation of a KK tower
coming from an honest extra dimension does have a local
continuum limit. We saw that the scale for the scattering of the light
KK modes was pushed above their $\Lambda_3$ through the exchange of 
additional massive gravitons and the massless graviphoton. The truncated
KK theory breaks down at the $\Lambda_3$ scale of the modes
whose tree-level scattering is affected by the truncation, namely
those halfway up the tower.

The truncated KK theory gives us much insight into what theory
spaces might come from an extra dimensional compactification.
We saw that exchange of gravitons on the second mass level
canceled strong four-point interactions of gravitons on the first mass
level. So it is unlikely that any compactification will have a KK tower
with the second lightest graviton {\it parametrically} heavier than the
first. More generally, we should not expect to see a parametrically large
mass gap between any set of modes.
It would be interesting to see what happens on AdS backgrounds, where
the strong coupling problem is ameliorated \cite{us1} 
and there is evidence that
at least one graviton mode can be made extremely light 
\cite{Karch:2000ct,Schwartz:2000ip}.
Also, note that the 
radion was not necessary for cancellation of the strongest
diagrams, and in fact couples with the same strength as the transverse
graviton modes.  
Although it contributes to cancel amplitudes which blow
up at $\Lambda_3 < \mv = \Lambda_{3/2}$, we 
have some freedom to manipulate
the radion in the effective theory. For example,
simply giving the radion a mass by hand will only affect
diagrams at the $\Lambda_2$ level applied to the lowest mode.
We did not explore the radion interactions
in depth, and in fact it may be necessary to perform
a wave-packet analysis to understand these sub-dominant amplitudes.

Incidentally, the significance of the $\Lambda_3$ scale is 
not at all clear. It seems to be more fundamental than the $\Lambda_5$
scale, which as was pointed out in Section \ref{secreview} is only
significant in the somewhat contrived Fierz-Pauli Lagrangian.
In contrast, $\Lambda_3$ appears as the
ultimate upper limit on the strong coupling scale of a theory with a single
massive graviton. It is even more intriguing 
that in a theory constructed to have
this strong coupling scale,
the natural size for all the operators in unitary gauge is
set by $\Lambda_3$ and dimensional analysis, a particularly clean
situation (see equation (4.40) of \cite{us1}).
Also, $\Lambda_3$, applied to the middle modes,
determines the scale where the truncated
KK theory breaks down. And we saw that
it is the strongest scale to which exchange of
transverse polarizations of the KK modes, and the radion, can contribute.
However, $\Lambda_3$ is not enough to use for the construction
of nearest neighbor discretizations which are guaranteed to be local in
the continuum limit.
Certainly, one important area for future investigation is the 
apparent coincidence of these various results.
 
Returning to the main theme of this paper,
it would be nice if the truncated KK theory had some exact symmetry
that the minimal discretization lacked.
The natural
candidate is the general coordinate reparameterization of $z$.
But because $z$ is the discretized dimension, this acts on KK towers
as the Virasoro algebra, which does not have any $N$-dimensional
representations. So the symmetry is broken in both theories.
In contrast,
the GC symmetries of the non-discretized dimensions
act faithfully on fields in the site basis.
Of course, these GC symmetries are a fake, broken in unitary gauge, 
and re-established formally
with the Goldstone bosons. But we cannot add
Goldstone bosons to restore the Virasoro symmetry in either model
because the fields belong to a truncated infinite dimensional multiplet.
The $U(1)$ symmetry for which the $\phi$ fields
are Goldstone bosons is also irrelevant. We saw in Section \ref{secu1}
that it cannot be made exact by extending Fierz-Pauli. Moreover, we know
the $\phi$ fields must exist in any discretization
because we need all the polarizations of the 4D massive gravitons
to get the five propagating modes of a massless graviton in 5D.
So the truncated KK theory does not seem to have any extra exact symmetries
at all.

In fact, 
there is nothing fundamentally
better about the truncated KK theory than the minimal discretization.
Both provide consistent low energy effective theories. 
While locality in the fifth dimension seems nice,
there is certainly no experimental evidence to support it.
And from the model building point of view, it is likely that there
are applications of gravity in theory space
for which locality is just irrelevant.
Recall that
the theory space technology was originally developed in gauge theory
to reproduce the phenomenology
of an extra dimension
\cite{Arkani-Hamed:2001ca,Hill:2000mu}, but it was soon seen
to be well-adapted 
for the construction of models with a naturally light ``little'' Higgs
\cite{Arkani-Hamed:2001nc}.
Similarly, 
gravitational theory spaces may produce applications
which have no extra dimensional interpretation at all.

Nevertheless, locality goes hand in hand with improved UV properties.
We have seen this already in Section \ref{secreview} where 
non-locality in the minimal discretization was traced to a
low cutoff in the effective theory. 
And certainly part of the motivation for trying to construct
gravitational dimensions comes from string theory, which
is both local and UV finite. While the nearest neighbor discretizations
have all the exact symmetries of the truncated KK theories, the latter
seem to have qualitatively superior UV properties. Now that we understand
the appropriate issues, we can work towards establishing a
more precise relation between apparent locality and
a higher cutoff. This may lead to a better understanding
of quantum gravity, and perhaps even a new class of UV completions.

\section*{Acknowledgements}
This work would not have been possible 
without the help of Nima Arkani-Hamed.
I would also like to thank Howard Georgi for his insight, especially
about the issues discussed in Section \ref{secu1}, Thomas Gregoire
for assisting with many of the Feynman diagram calculations and Paolo
Creminelli and Lisa Randall for helpful discussions.

\appendix

\section{Raising the Scale for a Single Massless Graviton 
\label{appproof}}
In this appendix it we show that a theory with a single
massive graviton must break down by $\Lambda_3$. We demonstrate
this by considering all possible extensions to the Fierz-Pauli
Lagrangian and calculating of all the
tree level amplitudes which must cancel. Because of subtleties
with a possible non-linearly realized $U(1)$ invariance,
this brute-force approach proves to be more convincing than
possible symmetry based arguments.

Start with the Fierz-Pauli Lagrangian:
\begin{equation}
\LFP
= \sqrt{g} R ( g ) + \frac{1}{4} ( [h^2] - [h]^2 )
\end{equation}
The strongest diagrams generated by this Lagrangian blow up
at $\Lambda_5$, but there are also diagrams which get strong at $\Lambda_4$
and $\Lambda_3$. More explicitly,
for
diagrams with four external lines, the $\Lambda_5$ is scalar
scattering through scalar exchange; $\Lambda_4$ is scalar
scattering; and $\Lambda_3$ is either 
vector scattering through scalar
exchange or scalar vector scattering. 
In general, 
vector exchange contributes at the same order as the corresponding
quartic vertex.

The only terms which may help cancel these tree-level scattering processes
are cubic and quartic in $h$:
%To raise this scale, we can consider adding the following
%due to scattering of the scalar modes of $h$. We can add the following 
%cubic and quartic extensions:
\begin{equation}
\Delta \cL = c_1 [h^3] + c_2 [h^2] [h] + c_3 [h]^3 + q_1 [h^4] + q_2 [h^2] [h^2]
+ q_3 [h^3] [h] + q_4 [h^2] [h]^2  + q_5 [h]^4
\end{equation}
It is not necessary to consider 
terms with space-time derivatives acting on $h$, because they 
contribute to amplitudes with different momentum dependence
than the ones we are trying to cancel.
%We will now show that
%this theory must be strongly coupled at $\Lambda_3$. 

Now, we want to study the interactions coming from these terms. 
As usual, we do this by introducing Goldstone bosons, 
via
$h \leadsto A +A^T + A^T A$
and invoking the equivalence theorem. 
We work at high energy, where
any gauge dependent mass the Goldstone may have is irrelevant, and
taken to be zero. 
This implies that
$[A]=[\Phi]=0$ if these fields correspond to external lines
({\it i.e.} on-shell, massless particles).
In particular, quartic terms which involve $[h]$ and cubic
terms which involve $[h]^2$ do not contribute,
at first order, to diagrams with four external longitudinal modes. 
So, $q_3 , q_4 , q_5$ and $c_3$ contribute
only above $\Lambda_3$ and we can ignore them. 

Thus, the interactions which may contribute up to fourth order
in $\LFP + \Delta \cL$ are:
\begin{multline}
\cL_A =( 1 + 6 c_1 ) [A^2 A^T] + ( - 1 + 4 c_2 ) [A] [A A^T] + 2 c_1 [A^3
] + 4 c_2 [A^2] [A]\\
+ 2 q_1 [A^4] + ( 6 c_1 + 8 q_1 ) [A^3 A^T] + ( 3 c_1 + 4 q_1 ) [A A A^T A^T]
+ ( \frac{1}{4} + 3 c_1 + 2 q_1 ) [A A^T A A^T]\\
+ ( - \frac{1}{4} + 2 c_2 + 4 q_2 ) [A A^T] [A A^T] + ( 2 c_2 + 8 q_2 ) [A^2]
[A A^T] + 4 q_2 [A^2] [A^2] \label{cla}
\end{multline}
The first processes we consider involve scalar exchange.
Replacing $A \rightarrow\Phi$ once for each of
the cubic terms in \eqref{cla} shows that the relevant interactions are:
%\be
%( 1 + 8 c_1 ) [\Phi^3] + ( - 1 + 8 c_2 ) [\Phi] [\Phi^2] \rightarrow
%\frac{1}{3} ( 1 + 24 c_1 + 16 c_2 ) [\Phi^3]
%\ee
\be
\cL_A \supset \phi ( ( 1 + 6 c_1 + 8 c_2 )_{} A_{\nu , \alpha , \mu} A_{\mu ,
\alpha , \nu} + ( - 1 + 8 c_2 ) A_{\alpha , \nu , \mu} A_{\alpha , \mu , \nu}
) \label{scalcurr}
\ee
We have integrated by parts to remove the derivatives from $\phi$.
Amplitudes coming from these vertices are strong at either 
$\Lambda_5$, if the external $A_\mu$ are longitudinally
polarized ({\it i.e.} $A_\mu \to \phi_{,\mu}$),
or at $\Lambda_3$, if the external
$A_\mu$ are transverse. Either way, the amplitude will be
proportional to the scalar current in \eqref{scalcurr} squared and
so \eqref{scalcurr} must exactly vanish. So $c_2 = \frac{1}{8}$
and $c_1 = -\frac{1}{3}$.

Next, we will look at the $\Lambda_3$ diagrams from the process 
$A A\rightarrow \Phi \Phi $, 
which get a contribution from vector exchange:
\begin{equation}
%
%  blob
%
\parbox{30mm}{
\begin{fmfgraph*}(60,50)
\fmfleft{p1,p2}
\fmfright{q1,q2}
\fmf{photon}{p1,p}
\fmf{photon}{p2,p}
\fmf{fermion}{p,q1}
\fmf{fermion}{p,q2}
\fmfblob{0.3w}{p}
\fmfv{l=$\overrightarrow{k}^\mu_2$,l.a=120,l.d=.05w}{p1}
\fmfv{l=$\overrightarrow{k}^\mu_1$,l.a=-120,l.d=.05w}{p2}
\fmfv{l=$\overrightarrow{k}^\mu_3$,l.a=60,l.d=.05w}{q1}
\fmfv{l=$\overrightarrow{k}^\mu_4$,l.a=-60,l.d=.05w}{q2}
\end{fmfgraph*} } =
%
%
%  quartic 
%
\parbox{20mm}{
\begin{fmfgraph*}(50,50)
\fmfleft{p1,p2}
\fmfright{q1,q2}
\fmf{photon}{p1,p}
\fmf{photon}{p2,p}
\fmf{fermion}{p,q1}
\fmf{fermion}{p,q2}
\end{fmfgraph*} } +
%
%  s channel
%
\parbox{30mm}{
\begin{fmfgraph*}(80,40)
\fmfleft{p1,p2}
\fmfright{q1,q2}
\fmf{photon}{p1,pl}
\fmf{photon}{p2,pl}
\fmf{photon,label=$A_\mu$}{pl,pr}
\fmf{fermion}{pr,q1}
\fmf{fermion}{pr,q2}
\end{fmfgraph*} }  +
%
% t-channel
%
\parbox{20mm}{
\begin{fmfgraph*}(40,70)
\fmfleft{p1,p2}
\fmfright{q1,q2}
\fmf{photon}{p1,pt}
\fmf{photon}{p2,pb}
\fmf{photon,label=$A_\mu$}{pt,pb}
\fmf{fermion}{pt,q1}
\fmf{fermion}{pb,q2}
%\fmfv{l=$p_2^A$,l.a=120,l.d=.05w}{p1}
%\fmfv{l=$p_1^A$,l.a=-120,l.d=.05w}{p2}
%\fmfv{l=$p_2^B$,l.a=60,l.d=.05w}{q1}
%\fmfv{l=$p_1^B$,l.a=-60,l.d=.05w}{q2}
\end{fmfgraph*} } +\quad
%
% u-channel
%
\parbox{20mm}{
\begin{fmfgraph*}(40,70)
\fmftop{p1,q1}
\fmfbottom{p2,q2}
\fmf{photon}{p1,pt}
\fmf{photon}{p2,pb}
\fmfforce{.5w,.75h}{pt}
\fmfforce{.5w,.25h}{pb}
\fmf{fermion}{pt,q2}
\fmf{fermion}{pb,q1}
\fmf{photon,label=$A_\mu$}{pt,pb}
%\fmfv{l=$p_1^A$,l.a=-120,l.d=.05w}{p1}
%\fmfv{l=$p_2^A$,l.a=120,l.d=.05w}{p2}
%\fmfv{l=$p_1^B$,l.a=-60,l.d=.05w}{q1}
%\fmfv{l=$p_2^B$,l.a=60,l.d=.05w}{q2}
\end{fmfgraph*} }  \nn
\end{equation}
The total amplitude, with the values for $c_1$ and $c_2$ we just
derived, is:
\begin{multline}
=( 4 q_1 - 1 ) [16 ( k_3 \varepsilon_1 ) ( k_3 \varepsilon_2 ) t u + 8 ( k_1
\varepsilon_2 ) ( k_3 \varepsilon_1 ) t^2 + 8 ( k_1 \varepsilon_2 ) ( k_3
\varepsilon_1 ) t u + 8 ( k_2 \varepsilon_1 ) ( k_3 \varepsilon_2 ) t u +
8 ( k_2 \varepsilon_1 ) ( k_3 \varepsilon_2 ) u^2]\\
+ ( 1 + 4 q_1 + 8 q_2 ) [8 ( k_1 \varepsilon_2 ) ( k_2 \varepsilon_1 ) ( t^2
+ u^2 )] + ( 1 + 8 q_1 + 16 q_2 ) [- 4 ( \varepsilon_1 \varepsilon_2 ) (
t^3 + u^3 )]\\
+ ( 3 + 4 q_1 + 24 q_2 ) [- 8 ( \varepsilon_1 \varepsilon_2 ) ( t^2 u + u^2
t ) + ( 3 + 16 q_1 + 32 q_2 ) [- 4 ( k_2 \varepsilon_1 ) 
( k_3 \varepsilon_2 ) ( t^2 + u^2 )] \\
+ ( 1 + 24 q_1 + 32 q_2 ) [- 4 ( k_3 \varepsilon_1 ) ( k_3 \varepsilon_2 ) (
t^2 + u^2 )] + ( 3 + 24 q_1 + 32 q_2 ) [4 ( k_1 \varepsilon_2 ) ( k_2
\varepsilon_1 ) t u]
\end{multline}
We have used
$k_4 = k_1 + k_2 - k_3$ and the
Mandelstam variables
$s\equiv k_1 k_2=k_3 k_4$, $t\equiv k_1 k_4=k_2 k_3$ 
and $u\equiv k_1 k_3=k_2 k_4$. 
The terms are grouped to illustrate that
no choice of $q_1$ and $q_2$ will make the entire
amplitude vanish. Therefore, there is no Lagrangian for
a single massive graviton which breaks down above $\Lambda_3$. 

\section{Cancellations in Truncated KK Theory \label{appfai}}
In this appendix we calculate the tree level scattering of the light
graviton modes in a Kaluza-Klein theory of 5D gravity. 
%We show that
%the strongest processes cancel through the sum of various contributions.
%An important result of this appendix is that it predicts the scale at
%which a truncated KK theory breaks down. 
Along the way, some subtle issues about setting up the calculation and
using the theory space formalism are addressed. In particular,
we now defend the choice of a circle, which initially
seems more complicated than the interval,
because of the extra graviphoton degree of freedom and the degenerate
spectrum. For a somewhat similar gauge
theory calculation the reader is referred to
\cite{Chivukula:2002ej,SekharChivukula:2001hz}.

For a general compactification, the KK wavefunctions $\chi^j ( z )$
depend on the the geometry and the boundary conditions. This lets the
Lagrangian be expressed, before the truncation, as the infinite sum
\begin{equation}
\cL = \sum o^2_{i j} G_i G_j + o^3_{i j k} G_i G_j G_k +
\cdots \label{olapL}
\end{equation} 
The $o^n$ are constants determined by overlap integrals of the $\chi^j$.
For example, if we compactify on a circle of radius $R$, then
$\chi_n ( z ) = e^{2 \pi i \frac{n z}{R}}$ and 
$o_{ij}=\int \chi_i \chi_j dz =\delta(i+j)$.

We now need a prescription for
going from the KK basis to the theory space basis. That is, we need to choose
discrete wavefunctions $g_i = c_{i j} G_j$. 
Of course, the obvious choice is a discrete Fourier sum, but we have to be
very careful. The essential requirement for a nice theory space
representation
is that
interactions which take place at fixed $z$ in the 5D theory 
(that is, those which do not depend on $\partial_z$) should
 transform to interactions which take place on
a single site $j$. That way, the $R_{4 D}$ part of $\cL$ \eqref{l5d4d}
will have a separate
general coordinate invariance for each site, and all the interactions of the
longitudinal modes will come from the $\partial_z$ part. This has
the advantage that all the interactions will be proportional to 
$\mpl^2 \partial_z^2 \sim\mpl^2 m^2$ 
instead of $\mpl^2 \partial^2 \sim \mpl^2 E^2 \gg \mpl^2 m^2$. 
The only way this will happen is if we can choose the discrete
wavefunctions to have all the same overlap integrals as the continuous
wavefunctions:
\begin{equation}
o_{i \cdots k} = \frac{1}{R} \int_0^R d z \chi_i \cdots \chi_k =
\frac{1}{N} \sum_n c_{i n} \cdots c_{k n} \label{overlaps}
\end{equation} 
For a general KK theory this is not possible. 
But if we compactify on a circle,
$\chi_n ( z ) = e^{2 \pi i \frac{n z}{R}}$, 
then $c_{a b} = e^{2 \pi i\frac{a b}{N}}$ 
satisfies \eqref{overlaps}. Of course, {\it any} KK theory has a theory
space representation, but in general the theory space
will not simplify the structure of the interactions. For example, it
is not possible to satisfy \eqref{overlaps} on an interval, and so
there will be strong interactions simply from the sites' Lagrangians.
While amplitudes involving these interactions must cancel
other amplitudes at the end of the day,
we get this cancellation for free if we choose to the circle where
\eqref{overlaps} holds.

\subsection{Circle Lagrangian}
The truncated KK theory on a circle has a degeneracy at each mass level.
We will need
the quadratic, cubic, and quartic parts of the KK Lagrangian
which come from \eqref{lkkmid}:
\begin{multline}
\cL \supset \mpl^2 \cL_K ( G_n G_{- n} )
-\mpl^2 m_1^2\Big[ \delta ( n + m ) m n ( [G_n G_m] - [G_n] [G_m] )\\
+ m n \delta ( m + n + p ) \{\frac{1}{2} [G_p] [G_m G_n] \} - 2 [G_m G_p
G_n] + [G_n] [G_p G_m] + [G_m] [G_p G_n] \}\\
+ m n \delta ( m + n + p + q ) \{- \frac{1}{4} [G_p G_q] [G_n G_m] - [G_n G_p]
[G_m G_q]
+ [G_n G_p G_m G_q] + 2 [G_n G_p G_q G_m] \}\Big] \label{L234}
\end{multline}
The delta functions
enforce KK number conservation, which is equivalent to momentum conservation
in the fifth dimension.
There are other cubic and quartic terms, but they are
irrelevant for the scattering processes we consider, and so we
omit them for clarity.
$\cL_K$ is the standard kinetic Lagrangian for a spin
two field, as shown in \eqref{hkin}, and has the same form for each
KK mode.

At this point, we could Fourier transform \eqref{L234}
to the site basis.
This is not hard, but 
the explicit form of the site Lagrangian is actually unnecessary
for the calculations we are interested in. 
We know that in unitary gauge, the Lagrangian will have
only one general coordinate invariance, under which all the $g_j$
transform as tensors. The $N-1$ broken GC symmetries correspond
to the massive gravitons, whose longitudinal modes produce the
strongest interactions in \eqref{L234}.
To study these interactions, we introduce Goldstone bosons on
each site in theory space.
That is, we replace each site metric:
\begin{equation}
g^j_{\mu \nu} \rightarrow \tilde{g}_{\mu \nu}^j = \frac{\partial
y_j^{\alpha}}{\partial x^{\mu}} \frac{\partial y_j^{\beta}}{\partial x^{\nu}}
 g^j_{\alpha \beta} ( y_j )
\end{equation} 
This restores $N$ copies of general coordinate invariance to the
Lagrangian. 

This brings up another subtle issue. The
transformation properties of the $y_a$ are not uniquely determined,
and different choices will actually
result in different interactions among
the Goldstones. Of course the scale of the strong interactions is
the same for all choices, but the calculations of individual amplitudes
may be very complicated if we do not introduce Goldstones in a
judicious way.
For now, we will make it so that the $\tilde{g}_{\mu \nu}^a$ are invariant
under all $N$ general coordinate transformations. This ensures
that the interactions among the Goldstones will respect a global
translation invariance around the circle.
The drawback of this choice is that 
the diagonal general coordinate invariance is now obscure and
the Lagrangian seems to depend explicitly
on $N$ sets of Goldstone bosons. 
Nevertheless, the Lagrangian must be independent
of some non-linear combination of these Goldstone fields.

The $y^\alpha$ are then expanded in terms of Goldstone bosons as usual 
$y^\alpha_j=x^\alpha+a^\alpha_j$ (and, as usual, we use lower-case
for the site basis).
The Goldstones are introduced in the site basis,
but immediately Fourier transformed: 
%\begin{equation}
%a_{\mu}^j = e^{- 2 \pi i n \frac{j}{N}} A^j_{\mu},
$a^j = e^{2\pi i \frac{j n}{N}} A^n$.
%\end{equation} 
Since in the KK basis
the masses are diagonal, the bilinear couplings of the Goldstones will be
diagonal as well. The quadratic terms in the KK basis are:
\begin{equation}
\mpl^2 ( n m_1 )^2 \Big\{[( G_n + A_n + A_n^T ) ( G_{- n} +
A_{- n} + A_{- n}^T )] - [G_n + 2 A_n] [G_{- n} + 2 A_{- n}] \Big\}
\end{equation} 
As with the single massive graviton case, the longitudinal modes of $A^n$ will
pick up kinetic terms from mixing with $G^n$. Thus, the fields have canonical
normalization as for a single graviton of mass $m_n = n m_1$.

The interactions we are interested in control the scattering of Goldstones.
So after introducing the Goldstones
in the site basis, we project out their interactions
as in Section \ref{secu1}
by setting $g_{\mu \nu}^j = \eta_{\mu \nu}$. 
In the matrix notation
(cf. \eqref{apa} and \eqref{amatnot}) this corresponds to:
\begin{equation}
g^j 
%\to (1+a^{j T})\tilde{g^j}(1+a^j)
\leadsto a^j + a^{j T} + a^{j T} a^j \label{introgb}
\end{equation} 
Because we have introduced the Goldstones on the sites, and they
are therefore summed over, this implies
\begin{equation}
G^n \to A^n + A^{n T} + A^{m T} A^{n-m}
\end{equation}
Thus, KK number will be preserved in Goldstone
scattering.

We have glossed over another subtle point.
Introducing Goldstone bosons on the sites is very different from
assigning a general coordinate invariance to each KK mode $G^n_{\mu\nu}$.
That would amount to replacing
$G^n \leadsto A^n + A^{n T} + A^{n T} A^{n}$
which leads to interactions which {\it violate} KK
number conservation. We are free to do this, and the physics will
be exactly the same
since both Lagrangians are the same in unitary
gauge, but it will be much harder to calculate. 
Not only will we have introduced KK number violating interactions,
but we will also have introduced interactions
into the $R_{4 D}$ part of the Lagrangian, which as we noted before, can
be canceled for free.

\subsection{Radion and Graviphoton}
The next step is to decide 
which of the massless fields have strong enough couplings
to contribute to scattering of the massive KK gravitons modes.
Since the radion, $r$, and the graviphoton, $V_\mu$, are massless,
they get normalized with
$\mpl$. Indeed, $V$'s kinetic term is already present in $\cL$
and the radion picks up a kinetic term from mixing with the massless
zero-mode graviton. Heuristically,
\be
\cL &=& \mpl^2 \sqrt{G_0} \{R ( G_0 ) e^r + 
( V_{\mu , \nu} - V_{\nu , \mu} )^2 \} +\cdots \\
&=& \mpl^2 G^0 \square G^0 + \mpl^2 r \square G^0 
+ \mpl^2 ( V_{\mu , \nu} - V_{\nu , \mu} )^2 +\cdots \nonumber
\ee
So canonical normalization is:
\be
V^c_{\mu} = \mpl V_{\mu} \quad\text{and}\quad r^c = \mpl r
\ee
Independent of the detailed tensor structure,
the couplings to the massive fields have the form:
\begin{equation}
\cL = 
\cdots + 
\frac{\partial^2}{\mpl} r^c A^c_n A^c_n + 
\frac{\partial^3}{\mpl m_n} V^c A^c_n A^c_n +
\frac{\partial^4}{\mpl m_n^2} r^c \Phi^c_n \Phi^c_n +
\frac{\partial^5}{\mpl m_n^3} V^c \Phi_n^c \Phi_n^c +
\cdots \label{rvint}
\end{equation} 
We will study below the strongest contributions to 
tree-level scattering
processes with each type of external lines. We can see from 
\eqref{rvint} that exchange of $r$ is a weaker process than
exchange of the Goldstone vector $A_\mu$, while $V_\mu$ and
$A_\mu$ exchange are the same strength. So, to first order, 
we can ignore $r$ but must include $V_\mu$.
We will therefore need the interactions which are linear in $V_{\mu}$ and
do not involve $r$ which were suppressed from the expression \eqref{lvintro}:
\begin{multline}
\cL_V = 
%- \frac{1}{4} ( V_{\mu , \nu} - V_{\nu , \mu} ) g^{\mu \nu} g^{\rho
%\sigma} ( V_{\rho , \sigma} - V_{\sigma , \rho} ) \\
%+ 
g^{\alpha \beta} g^{\mu \nu} g^{\gamma \delta} V_{\beta} ( 2
g_{\alpha \mu, z} g_{\nu \gamma , \delta} 
- g_{\alpha \mu,z} g_{\gamma \delta ,\nu} 
- g_{\gamma \delta,z} g_{\alpha \mu , \nu} 
+ \frac{1}{2} g_{\gamma \delta,z} g_{\mu \nu , \alpha} 
- \frac{3}{2} g_{\gamma \mu,z} g_{\delta \nu , \alpha} 
+ g_{\gamma \mu,z} g_{\nu \alpha , \delta} ) \\
+ g^{\mu \nu} g^{\alpha \beta} 
( V_{\beta , \alpha} g_{\mu \nu,z} - V_{\beta , \mu} g_{\alpha \nu,z} ) 
+ g^{\mu \nu} V_{\beta} ( 2 g_{\mu \nu , \alpha,z} 
- 2 g_{\alpha \mu , \nu,z} ) + \cdots \label{lv}
\end{multline}
All of the couplings
that we will need below come from the following terms in the 
KK decomposition of \eqref{lv}:
\begin{equation}
\cL \supset - i \mpl^2 m_1 n \delta ( n + m ) V_{\beta} ( 2 G^m_{\mu \nu}
G^n_{\mu \nu , \beta} + G^m_{\mu \nu , \beta} G^n_{\mu \nu} - 2 G^m_{\mu \nu}
G^n_{\mu \beta , \nu} ) \label{Vint}
\end{equation} 

Now let us come back to one of the subtle issues mentioned above.
In addition to the radion and the KK gauge boson, we must contend with
the zero mode Goldstone boson, $A_{\mu}^0$. It does not have a kinetic
term, and does not pick one up by mixing. But it does have interactions, with
the Lagrangian in its current form. 
The only reason we have this mode at all, is because
when we introduced
the Goldstone bosons in \eqref{introgb} 
we included one set for each of the $N$ sites,
even though the Lagrangian has only $N - 1$ broken coordinate invariances.
The interactions of $A_\mu^0$ exist
 because the preserved symmetry is not the one
under which all the fields on the sites
transform nicely; it is the one where all the
KK modes transform nicely.
Suppose we had chosen the transformations of
the Goldstone bosons so that each $\tilde{g}_{\mu \nu}^j$ were covariant
under changes of a single coordinate $y^{s}$ and invariant under all the
others. Then, 
$\tilde{g}_{\mu\nu}^{s} = g_{\mu \nu}^s$
and
$\tilde{g}_{\mu\nu}^j = y^{\alpha}_{j,\mu} 
y^{\beta}_{j,\nu} g_{\alpha \beta}^j$ 
for $j \neq s$.
Then there would be only $N-1$ sets of
Goldstones and we would not have the peculiar $A_{\mu}^0$ field. The
Lagrangian would be the same, but with $a^s_\mu = 0.$ In terms of the KK
modes of the Goldstones, this implies
\begin{equation}
A^0_{\mu} = - e^{2 \pi i \frac{s n}{N}} A^n_{\mu} \label{a0sub}
\end{equation} 
For example, if we we take $s = 0$ then $A^0_{\mu} \rightarrow -
A_{\mu}^1 - A_{\mu}^2 - \cdots $. Now, there will be many KK number violating
vertices. In particular, all the heavy KK modes would contribute to
scattering of the low modes. But when we sum over all diagrams, KK number
should not be violated. It would be nice if we could just use the residual
general coordinate invariance to set $A_{\mu}^0 = 0$, but it not clear that
this is consistent. There is no simple way of introducing Goldstones so that
$A_{\mu}^0$ does not appear at all. As it turns out, if we calculate the
scattering from terms generated by the substitution \eqref{a0sub} 
into the cubic and
quartic vertices involving the $A^0_{\mu}$, everything vanishes. This implies
that we are probably free to just set $A_{\mu}^0 = 0$.

\subsection{Goldstone Interactions}
From now on, we use the notation $1 \equiv A^1_{\mu , \nu}$, 
where $A_{\mu}^1$
is the first KK mode of the vector of Goldstone bosons introduced. 
Scattering of the the two modes in the first massive level
involves exchange of the two modes on the second massive level.
So we need $N \geq 5$ to see a non-trivial
cancellation. Therefore, we take $N=5$.
Note that taking $N > 5$ will not change the relevant vertices, 
as these get no contribution from higher KK modes.
So we consider the lightest five modes $0,1,2,-1\equiv \overline{1}$ and
$-2\equiv \overline{2}$, and restrict to diagrams involving only external
$1$ and $\overline{1}$.
We will also immediately go to 
canonical normalization
$A^n_{\mu} \rightarrow \mpl m_n A^n_{\mu}$ 
and $\Phi^n \rightarrow \mpl m_n^2 \Phi^n$.
Now, we present the relevant interactions among these fields.

We start with the cubic vertices. 
After a lengthy calculation we can isolate
the following terms relevant for diagrams with external $1$ and $\overline{1}$:

\begin{equation}
\cL_C \supset 8 [1 1^T{\overline 2}] - 4 [1^T
1{\overline 2}] + 12 [1 1{\overline 2}] + 8 [1^T
1^T{\overline 2}] - 14 [11] [{\overline 2}] + 2 [1^T 1]
[{\overline 2}] \nonumber
\end{equation} 
\begin{equation}
- 16 [1] [1\overline{2}] - 8 [1] [1^T{\overline 2}] + 12
[\overline{2}] [1] [1] + c . c . \label{lc12}
\end{equation}
It is interesting to note that we can perform a field redefinition to remove
all the cubic terms:
\begin{equation}
2_{\mu} \rightarrow 2_{\mu} + \frac{1}{8} 1_{\beta} 1_{\mu , \beta}
\end{equation}
Note that this substitution preserves KK number conservation.
This is an example of the type of non-linear transformation we were
wary of in Section \ref{secu1}. But it does not actually
make the calculations of this appendix any easier, so we leave
\eqref{lc12} as it is.

Next, the couplings of the graviphoton can be written as (from \eqref{Vint}):
\begin{equation}
\cL \supset 
4 i V_{\beta} ( 1_{\mu , \nu} \overline{1}_{\beta , \mu , \nu} -
\overline{1}_{\mu , \nu} 1_{\beta , \mu , \nu} ) \label{lcvab}
\end{equation} 
We have already dropped everything which vanishes when the Goldstones,
which always appear on shell in $V$-exchange diagrams, are massless. The
interactions in \eqref{lcvab}
contribute to $11$ scattering at the order $\Lambda_3$,
which is the same order as the quartic vertices and as $2$ exchange.

Next, we look at the quartic vertices with  $1$ and $\overline{1}$:
\begin{multline}
\cL_Q \supset - 8 [{\overline 1}{\overline 1} 11] - 4
[1^T{\overline 1}{\overline 1} 1] + 4 [{\overline 1}
{\overline 1} 1^T 1] - 4 [{\overline 1}^T 11 {\overline 1}]
+ 4 [{\overline 1}^T{\overline 1} 11] \nn\\
- 4 [{\overline 1}^T{\overline 1}^T 11] + 2 [{\overline 1}^T 1^T 
1{\overline 1}] + 2 [1^T{\overline 1}^T {\overline 1} 1]
- 4 [{\overline 1} 1{\overline 1} 1] - 4 [{\overline 1}^T
1^T{\overline 1} 1] - 4 [{\overline 1}^T 1{\overline 1}
1] - 4 [1^T{\overline 1} 1{\overline 1}] \nn\\
+ 6 [11] [{\overline 1}{\overline 1}] - [11] [{\overline 1}
{\overline 1}^T] - [11^T] [{\overline 1}{\overline 1}]
+ 4 [1{\overline 1}] [1{\overline 1}] + 4 [1{\overline 1}] [1{\overline 1}^T]\\
\end{multline}
There are other terms involving $[1]$ and $[\overline{1}]$. These will not
contribute when the external lines are on shell,
so we have not displayed them here.

To calculate the interactions, it is easiest to project out the real and
imaginary parts of the KK fields. This diagonalizes the kinetic terms of the
physical fields. To account for the normalization as well, we make the
substitution $1 = \frac{1}{\sqrt{8}} ( A + i B )$ and $2 = \frac{1}{2
\sqrt{8}} ( C + i D )$. 
Then the entire normalized Lagrangian we will need is:
\begin{multline}
\cL \supset - \frac{3}{16} [A^4] - \frac{1}{8} [A^T A^3] - \frac{1}{16} [A^T A^T
A^2] + \frac{5}{32} [A^2]^2 + \frac{1}{32} [A^2] [A A^T]\\
- \frac{3}{16} [B^4] - \frac{1}{8} [B^T B^3] - \frac{1}{16} [B^T B^T B^2] +
\frac{5}{32} [B^2]^2 + \frac{1}{32} [B^2] [B B^T] \\
- \frac{1}{8} [A B A B] + \frac{1}{8} [A^T B A B] + \frac{1}{8} [B^T A B A] +
\frac{1}{8} [A^T B^T A B]\\
- \frac{1}{4} [A^2 B^2] - \frac{3}{8} [A^T A B B] - \frac{3}{8} [B^T B A A] -
\frac{3}{16} [A^T B^T B A] - \frac{3}{16} [B^T A^T A B]\\
+ \frac{1}{8} [B^T A A B] + \frac{1}{8} [A^T B B A] + \frac{1}{8} [A^T A^T B B ]
- \frac{1}{16} [B^2] [A^2] + \frac{3}{8} [A B] [A B]\\
 + \frac{3}{32} [A^2] [BB^T] + \frac{3}{32} [B^2] [A A^T] 
- \frac{1}{8} [A B] [A B^T] \\
%\end{multline}
%\begin{multline}
+ \frac{1}{\sqrt{8}} [A A^T C] + \frac{1}{\sqrt{8}} [A^T A^T C] -
\frac{1}{\sqrt{8}} [A A] [C]\\
- \frac{1}{\sqrt{8}} [B B^T C] - \frac{1}{\sqrt{8}} [B^T B^T C] +
\frac{1}{\sqrt{8}} [B B] [C]  \\
+ \frac{1}{\sqrt{8}} [A B^T D] + \frac{1}{\sqrt{8}} [B A^T D] +
\frac{1}{\sqrt{8}} [A^T B^T D] + \frac{1}{\sqrt{8}} [B^T A^T D] -
\frac{1}{\sqrt{2}} [A B] [D]\\
- \frac{1}{2} [A^T A] + \frac{1}{2} [A^2] - \frac{1}{2} [B^T B] + \frac{1}{2}
[B^2] - \frac{1}{2} [C^T C] + \frac{1}{2} [C^2] - \frac{1}{2} [D^T D] +
\frac{1}{2} [D^2]\\
-\frac{1}{4} ( V_{\mu , \nu} - V_{\nu , \mu} )^2 + V_{\beta} ( A_{\mu ,
\nu} B_{\beta , \mu , \nu} - B_{\mu , \nu} A_{\beta , \mu , \nu} ) \label{LAB}\\
\end{multline}

\subsection{Test the Lagrangian \label{sectest}}
To test the Lagrangian, we will look at some characteristic processes. Recall
that the strength of a vertex for involving a graviton of mass $m_g$
is given by 
$\Lambda_{\lambda} =(m_g^{\lambda-1} \mpl)^{1/\lambda}$
where $\lambda$ is given
by \eqref{lamform}. 
The strongest processes involve scalar exchange. And the vector
exchange diagrams have the same strength as the diagrams coming from the
quartic vertices as well as the diagrams with graviphoton exchange. If we
separate out scalar exchange, which must cancel by itself, then the strength
of a process is determined by the external lines. Since this is the case, we
do not have even to project out the scalar in the external lines, we
can just leave it as the longitudinal polarization of the vector field. We
only insist that the polarization of the eternal vector field satisfies
$\varepsilon_{\mu , \mu} = 0$.

The masses of the Goldstone bosons are gauge dependent,
but we will only be concerned with the lowest order tree level
effects, and so we take all the Goldstones to be massless. The corrections,
of order $m^2 / k^2$ will only contribute to higher order processes which we
will ignore. In practice, this means dropping terms which contain $[A] , [B]
, A_{\mu , \mu} , B_{\mu , \mu} , A_{\mu , \nu , \nu}$ or $B_{\mu , \nu , \nu
}$. We have already done this for the Lagrangian \eqref{LAB} but more
simplifications come about for particular process.

The cubic terms' contribution to $C$ and $D$ exchange can be written as:
\be
\cL \supset &-& \frac{1}{\sqrt{2}} C_{\mu} ( A_{\alpha , \beta} A_{\mu ,
\alpha , \beta} - A_{\alpha , \beta} A_{\beta , \alpha . \mu} - B_{\alpha ,
\beta} B_{\mu , \alpha , \beta} + B_{\alpha , \beta} B_{\beta , \alpha , \mu}
) \label{lcc} \\
&-& \frac{1}{\sqrt{2}} D_{\mu} ( A_{\nu , \alpha} B_{\mu , \alpha , \nu} +
A_{\mu , \alpha , \beta} B_{\beta , \alpha} - A_{\alpha , \beta , \mu}
B_{\beta , \alpha} - A_{\alpha , \beta} B_{\beta , \alpha , \mu} ) \label{lcd}
\ee 
It is easy to see that for scalar exchange $C_{\mu} \rightarrow \phi^C_{,
\mu}$ and $D_{\mu} \rightarrow \phi_{, \mu}^D$ the above terms vanish
after integrating by parts. This means that all the scalar exchange
processes, which are stronger than the corresponding 
vector exchange processes,
vanish.

Next, consider the process $\phi^A \phi^A \rightarrow \phi^B
\phi^B$. This contributes at the scale $\Lambda_4$. If we make the
substitution $A_{\mu} \rightarrow \phi^A_{, \mu}$ and $B_{\mu}
\rightarrow \phi^B_{, \mu}$ into \eqref{lcc} and \eqref{lcd} we see that 
there is no contribution from vector $C$ and $D$ exchange:
\vskip 2mm
%
%   s channel
%
\begin{equation}
\parbox{40mm}{
\begin{fmfgraph*}(100,40)
\fmfleft{p1,p2}
\fmfright{q1,q2}
\fmf{fermion,label=$\phi^A$,label.side=right}{p1,pl}
\fmf{fermion,label=$\phi^A$,label.side=left}{p2,pl}
\fmf{photon,label=$C,,D$}{pl,pr}
\fmf{fermion,label=$\phi^B$,label.side=right}{pr,q1}
\fmf{fermion,label=$\phi^B$,label.side=left}{pr,q2}
\end{fmfgraph*} } \ = 0
\end{equation}
The quartic terms are:
\begin{equation}
\cL \supset - [\Phi^A \Phi^A \Phi^B \Phi^B] + \frac{1}{4} [\Phi^A \Phi^B
\Phi^A \Phi^B] + \frac{1}{4} [\Phi^A \Phi^B] [\Phi^A \Phi^B] +
\frac{1}{8} [\Phi^A \Phi^A] [\Phi^B \Phi^B]
\end{equation}
This does not vanish by integration by parts. If we define the Mandelstam
variables $s = p^A_1 \cdot p^A_2$, $t = p_1^A \cdot p_1^B$ and $u = p_1^A
\cdot p_2^B$ then this contributes:
%
%  quartic 
%
\begin{equation}
\parbox{35mm}{
\begin{fmfgraph*}(70,40)
\fmfleft{p1,p2}
\fmfright{q1,q2}
\fmf{fermion,label.side=left}{p1,p}
\fmf{fermion,label.side=right}{p2,p}
\fmf{fermion,label.side=left}{p,q1}
\fmf{fermion,label.side=right}{p,q2}
\fmfv{l=$p_2^A$,l.a=120,l.d=.05w}{p1}
\fmfv{l=$p_1^A$,l.a=-120,l.d=.05w}{p2}
\fmfv{l=$p_2^B$,l.a=60,l.d=.05w}{q1}
\fmfv{l=$p_1^B$,l.a=-60,l.d=.05w}{q2}
\end{fmfgraph*} } \ = 
- 2 s^2 ( t^2 + u^2 ) + t^2 u^2 + \frac{1}{2} ( t^4 + u^4 ) + \frac{1}{2} s^4
\neq 0
\end{equation} 
However, there is also a contribution from the graviphoton:
\begin{equation}
\cL \supset V_{\beta} ( \phi^A_{, \mu , \nu} \phi^B_{, \beta , \mu , \nu} -
\phi^B_{, \mu , \nu} \phi^A_{,\beta , \mu , \nu} )
\end{equation} 
This contributes through the $t$- and $u$- channels:
\begin{equation}
%
% t-channel
%
\parbox{20mm}{
\begin{fmfgraph*}(40,60)
\fmfleft{p1,p2}
\fmfright{q1,q2}
\fmf{fermion}{p1,pt}
\fmf{fermion}{p2,pb}
\fmf{photon,label=$V$}{pt,pb}
\fmf{fermion}{pt,q1}
\fmf{fermion}{pb,q2}
\fmfv{l=$p_2^A$,l.a=120,l.d=.05w}{p1}
\fmfv{l=$p_1^A$,l.a=-120,l.d=.05w}{p2}
\fmfv{l=$p_2^B$,l.a=60,l.d=.05w}{q1}
\fmfv{l=$p_1^B$,l.a=-60,l.d=.05w}{q2}
\end{fmfgraph*} } +\quad
%
% u-channel
%
\parbox{20mm}{
\begin{fmfgraph*}(40,70)
\fmftop{p1,q1}
\fmfbottom{p2,q2}
\fmf{fermion}{p1,pt}
\fmf{fermion}{p2,pb}
\fmfforce{.5w,.75h}{pt}
\fmfforce{.5w,.25h}{pb}
\fmf{fermion}{pt,q2}
\fmf{fermion}{pb,q1}
\fmf{photon,label=$V$}{pt,pb}
\fmfv{l=$p_1^A$,l.a=-120,l.d=.05w}{p1}
\fmfv{l=$p_2^A$,l.a=120,l.d=.05w}{p2}
\fmfv{l=$p_1^B$,l.a=-60,l.d=.05w}{q1}
\fmfv{l=$p_2^B$,l.a=60,l.d=.05w}{q2}
\end{fmfgraph*} } 
\ = 
- \left\{\frac{t^4 ( 2 s + 2 u )}{- 2 t} + \frac{u^4 ( 2 s + 2 t )}{- 2 u}
\right\} 
\end{equation} 
If we apply the relation $s - t - u = 0$ the quartic and exchange
contributions exactly cancel.

Note that the $C$ and $D$ fields do not contribute at  either $\Lambda_5$
(scalar scattering through scalar exchange) or at  $\Lambda_4$
(scalar scattering through vector exchange). So even if we
truncated the theory at the level at the first massive mode, the strong
coupling scale would already be $(m_1^2 \mpl)^{1/3}$.
However, to see the cancellation of the remaining diagrams, we need to include
the effects of the heavier fields.

It is straightforward to work through remaining scattering processes. The
computations are more involved, but at tree-level all the amplitudes
involving the Lagrangian \eqref{LAB} are exactly zero.

\end{fmffile}

\bibliographystyle{JHEP}

\end{document}